\documentstyle[amsfonts,12pt]{article}
\title{}\date{}

\title{Exact Kink Solitons in Skyrme Crystals}

\author{Shouxin Chen\\Institute of Contemporary Mathematics\\School of Mathematics\\Henan University\\
Kaifeng, Henan 475004, PR China\\\\Yijun Li\\School of Mathematics\\Henan University\\
Kaifeng, Henan 475004, PR China\\\\Yisong Yang\\Department of Mathematics\\Polytechnic Institute of New York University\\Brooklyn, New York 11201, USA}

\newcommand{\bfZ}{{\Bbb Z}}

\def\XXint#1#2#3{{\setbox0=\hbox{$#1{#2#3}{\int}$}
 \vcenter{\hbox{$#2#3$}}\kern-.5\wd0}}

\newtheorem{oldtheorem}{Theorem}[section]
\newtheorem{oldassertion}[oldtheorem]{Assertion}
\newtheorem{oldproposition}[oldtheorem]{Proposition}
\newtheorem{oldremark}[oldtheorem]{Remark}
\newtheorem{oldlemma}[oldtheorem]{Lemma}
\newtheorem{olddefinition}[oldtheorem]{Definition}
\newtheorem{oldclaim}[oldtheorem]{Claim}
\newtheorem{oldcorollary}[oldtheorem]{Corollary}

\newbox\qedbox
                        {\hfill\penalty10000\copy\qedbox\par\medskip}

\newcommand{\dd}{\mbox{d}}
\newcommand{\ee}{\end{equation}}
\newcommand{\be}{\begin{equation}}\newcommand{\bea}{\begin{eqnarray}}
\newcommand{\eea}{\end{eqnarray}}

\newcommand{\pa}{\partial}

\newcommand{\nn}{\nonumber}

\newcommand{\lm}{\lambda}

\begin{document}
\maketitle
\begin{abstract}
We present an explicit integration of the kink soliton equation obtained in a recent interesting study of the classical Skyrme model where the field configurations are of
a generalized hedgehog form which is of a domain-wall type. We also show that in such a reduced one-dimensional setting the first-order and second-order equations are equivalent.
Consequently, in such a context, all finite-energy solitons are BPS type and precisely known.

\medskip

\end{abstract} 

\maketitle

\medskip 
\medskip 

The well-known study of Derrick \cite{D} shows that for a wide class of nonlinear wave equations there exist no stable time-independent solutions of finite energy due
to the conformal structure of the standard Euclidean space. In order to overcome such a difficulty one may extend the theory to contain gauge fields so that stable time-independent
solutions of finite energy, known as vortices, monopoles, and instantons, exist in two, three, and four spatial dimensions \cite{JT,MS,R,Ybook}. Alternatively, Skyrme \cite{S}
showed that it
is possible to introduce higher-order nonlinear terms involving derivatives, instead of gauge fields, to overcome the conformal structure problem of the Derrick type so that
continuously behaved, topologically characterized, spatially concentrated, and time-independent solutions of finite energy are allowed to exist.
However, unlike in the gauge-field models where BPS type reductions, after Bogomol'nyi \cite{B} and Prasad and Sommerfield \cite{PS}, are present to enable explicit constructions of  solutions \cite{AD,W,R,A,NS},
no explicit constructions for the Skyrme solitons are available in literature due to the fact that even though a BPS topological lower bound may be derived but it cannot be saturated
by nontrivial field configurations \cite{M,MR}, although some partial progress has been developed for an existence theory based on nonlinear functional analysis
\cite{E,L,LY}. For a recent work on the BPS lower bound and its saturation for the Skyrme models in rather general settings, see Adam and Wereszczynski \cite{AW},
which also refers to other related developments of the subject.

In an interesting study of Canfora \cite{C1} developed from some recent earlier work \cite{C2,C3}, a static ansatz periodic in two planar coordinate directions is used to 
reduce the Skyrme energy to that of a kink or domain wall given in terms of a third spatial coordinate direction perpendicular to the periodic plane. The second-order governing
equation takes a complicated form but it may be solved by a much simpler first-order equation whose solutions interpolate adjacent minima of the one-dimensional Skyrme energy density.
The importance of the work of Canfora is that it makes an explicit construction of kink-like solutions in the classical Skyrme model possible, which prompts our investigation.
Specifically, in the present note, we follow \cite{C1} to obtain an explicit integration of the first-order equation found in \cite{C1}. Moreover, we prove that the first-order equation of Canfora \cite{C1} and the 
second-order equation of the one-dimensional kink energy of Canfora \cite{C1} are actually equivalent under the finite-energy condition. Therefore all finite-energy solutions of the
Skyrme model within the Canfora kink ansatz \cite{C1} are obtained explicitly.

In the rest of this note, we first recall the one-dimensional reduction of Canfora \cite{C1} (see also \cite{C2,C3}) of the Skyrme model. We next show that in the context of
finite-energy solutions the first-order and second-order equations of Canfora \cite{C1} are equivalent. We then carry out an explicit integration of the first-order equation.

Let $\mu,\nu$ denote the Minkowski spacetime indices with metric $\mbox{diag}\{-1,1,1,1\}$. In normalized form the Skyrme action density reads
\be \label{1}
{\cal L}=\frac14\mbox{Tr}\left(R^\mu R_\mu+\frac\lm8 F_{\mu\nu}F^{\mu\nu}\right),
\ee
where $\lm>0$ is a coupling parameter, $R_\mu=U^\dagger \pa_\mu U=R^i_\mu t_i$ for an $SU(2)$-valued map $U$, $t_i$ ($i=1,2,3$) are generators of the Lie algebra of $SU(2)$,
and $F_{\mu\nu}=[R_\mu,R_\nu]$, so that the associated Euler--Lagrange equations (or the Skyrme field equations \cite{C1}) are
\be \label{2}
\pa^\mu R_\mu+\frac\lm4\,\pa^\mu[R^\nu,F_{\mu\nu}]=0.
\ee
When $U$ is represented through the expression
\be \label{3}
U=Y^0{\bf 1}+Y^i t_i,\quad (Y^0)^2+Y^iY_i=1,
\ee
where $\bf 1$ denotes the $2\times2$ identity matrix and $Y^0,Y^i$ are scalar functions over spacetime, then the static hedgehog-like ansatz adopted in \cite{C1,C2,C3} takes the form
\be
Y^0=\cos\alpha,\quad Y^i=\hat{n}^i\sin\alpha,
\ee 
so that $\hat{n}^i$ is a unit $3$-vector satisfying
\be \label{5}
\hat{n}^1=\cos\Theta,\quad \hat{n}^2=\sin\Theta,\quad\hat{n}^3=0,\quad \pa_\mu\alpha \pa^\mu\Theta=0.
\ee
In the static case, the last condition in (\ref{5}) indicates that the vectors $\nabla\alpha$ and $\nabla\Theta$ stay mutually perpendicular.
For convenience, use the notation $t=x^0,x=x^1,y=x^2,z=x^3$ and take \cite{C1} the further ansatz
\be 
\alpha=\alpha(x),\quad \Theta=\Theta(y,z)=\frac1l(n_2 y+n_3 z),\quad n_2,n_3\in\bfZ.
\ee
Then the Skyrme equations (\ref{2}) become a single one \cite{C1},
\be \label{7}
\alpha''=\frac1{2L^2}\sin(2\alpha)-\frac\lm{L^2}\left(\alpha''\sin^2\alpha+\frac{(\alpha')^2}2\sin(2\alpha)\right),
\ee
with $\alpha'=\dd\alpha/\dd x$, $L=l/\sqrt{n_2^2+n_3^2}$, and the reduced energy (Hamiltonian) density
\be \label{8}
{\cal H}(\alpha)\equiv T_{tt}=\frac12\left((\alpha')^2+\frac1{L^2}\sin^2\alpha+\frac\lm{L^2}(\alpha')^2\sin^2\alpha\right),
\ee
associated with the action density (\ref{1}). Vanishing of the energy density (\ref{8}) spells out that the ground states are given by
\be \label{9}
\alpha=n\pi,\quad n\in\bfZ,
\ee
at $x=\pm\infty$ and one is to find solutions to (\ref{7}) subject to the boundary condition (\ref{9}) at $x=\pm\infty$, which amounts to solving 
a two-point boundary value problem which seems difficult.

Since the $Y^3$-component is absent, the Skyrme charge, or the topological degree, also referred to as the baryon number, of the map $U$ stated in (\ref{3}) is zero. Such a feature limits, perhaps, the possible applications of the setting to particle physics. Nevertheless, Canfora \cite{C1} identifies 
a nontrivial kink charge $Q$ given by
\bea\label{10}
Q&=&\int_{-\infty}^\infty\frac1L\sin\alpha\left(1+\frac\lm L^2\sin^2\alpha\right)^{\frac12}\alpha'\,\dd x\nn\\
&=&\pm\left\{\frac1{2L}\left(1+\sqrt{1+\frac{2\lm^2}L}\right)+\frac1{\lm\sqrt{L}}\left(1+\frac{\lm^2}L\right)\arcsin\frac{\lm}{\sqrt{L+\lm^2}}\right\},
\eea
with a toroidal topology characterization, so that he obtains through a BPS trick \cite{B,PS} the following expression for the total energy per unit area of the $yz$-domain:
\bea \label{11}
E&=&\int_{-\infty}^\infty{\cal H}(\alpha)\,\dd x=\int_{-\infty}^\infty \frac12\left((\alpha')^2+\frac1{L^2}\sin^2\alpha+\frac\lm{L^2}(\alpha')^2\sin^2\alpha\right)\,\dd x\nn\\
&=&\int_{-\infty}^\infty\left(1+\frac\lm L^2\sin^2\alpha\right)\left(\alpha'\pm\left[1+\frac\lm{L^2}\sin^2\alpha\right]^{-\frac12}\frac{\sin\alpha}L\right)^2\,\dd x\mp Q\nn\\
&\geq&|Q|,
\eea
for $|Q|=\mp Q$, which leads him to arrive at the conclusion that the energy lower bound in (\ref{11}) is attained when $\alpha$ satisfies the first-order equation
\be\label{12}
\alpha'\pm\left(1+\frac\lm{L^2}\sin^2\alpha\right)^{-\frac12}\frac{\sin\alpha}L=0.
\ee

It is straightforward to examine that (\ref{12}) implies (\ref{7}). We now prove the converse: any finite-energy solution of (\ref{7}) satisfies (\ref{12}) as well.
That is, the first-order equation (\ref{12}) and the second-order equation (\ref{7}) are actually equivalent.

In fact, for convenience, we rewrite (\ref{7}) and (\ref{12}) as
\be \label{13}
\alpha''=\frac{\sin(2\alpha)(1-\lm(\alpha')^2)}{2(L^2+\lm\sin^2\alpha)},
\ee
and
\be\label{14}
\alpha'\pm\frac{\sin\alpha}{\sqrt{L^2+\lm\sin^2\alpha}}=0,
\ee
respectively. We proceed as follows.

Let $\alpha$ be a finite-energy solution of (\ref{13}) and set
\be\label{15}
P_\pm={\sqrt{L^2+\lm\sin^2\alpha}}\,\,\alpha'\pm\sin\alpha.
\ee
Then, in view of (\ref{13}), we have
\be\label{16}
P_\pm'=\pm\frac{\cos\alpha}{\sqrt{L^2+\lm\sin^2\alpha}} P_\pm.
\ee

Therefore, if there is some $x_0\in(-\infty,\infty)$ such that $P_+(x_0)=0$ or $P_-(x_0)=0$, then applying the uniqueness theorem for the initial value problem of an ordinary
differential equation we obtain $P_+\equiv0$ or $P_-\equiv0$, which implies that $\alpha$ must satisfy one of the equations stated in (\ref{14}).

Observe that the elementary inequality $(a\pm b)^2\leq 2(a^2+b^2)$ and the finite-energy condition lead us to
\be \label{17}
\int_{-\infty}^\infty P_\pm^2(x)\,\dd x<\infty.
\ee
Inserting (\ref{17}) into (\ref{16}), we also have
\be\label{18}
\int_{-\infty}^\infty (P_\pm'(x))^2\,\dd x<\infty.
\ee
Combining (\ref{17}) and (\ref{18}) we conclude with
\be \label{19}
\lim_{|x|\to\infty}P_\pm(x)=0.
\ee
Furthermore, using (\ref{16}) again, we get
\be\label{20}
(P_+P_-)'=0,\quad x\in(-\infty,\infty).
\ee
In view of (\ref{19}) and (\ref{20}), we deduce 
\be 
(P_+P_-)(x)=0,\quad x\in(-\infty,\infty).
\ee
That is, for any $x\in(-\infty,\infty)$, either $P_+(x)=0$ or $P_-(x)=0$, which establishes $P_+\equiv0$ or $P_-\equiv0$ as anticipated, as argued earlier.

Since (\ref{13}) and (\ref{14}) are equivalent, we can now concentrate on (\ref{14}), subject to the two-point boundary value condition
\be \label{22}
\alpha(-\infty)=m\pi,\quad \alpha(\infty)=n\pi.
\ee

Applying the uniqueness theorem for the initial value problem of an ordinary
differential equation to (\ref{14}) again we see that if there is some $x_0\in (-\infty,\infty)$ such that $\alpha(x_0)=k\pi$ for some $k\in\bfZ$ then
$\alpha(x)=k\pi$ for all $x\in(-\infty,\infty)$ which trivializes the solution. From now on we only consider the nontrivial situation. Consequently we may assume $\alpha'(x)\neq0$
for all $x$ by virtue of (\ref{14}).

If $\alpha'(x)>0$ for all $x$, then it is necessary to have $m<n$ in (\ref{22}). Since $\alpha(x)$ can never attain a value which is an integer multiple of $\pi$, we must have
$m=n-1$ (the neighboring situation recognized already in \cite{C1}). Thus $n-1$ or $n$ must be even. For definiteness we assume $n-1$ is even. Hence $\sin\alpha$ stays positive
in $(-\infty,\infty)$ and consistency indicates that we encounter the minus sign situation given in (\ref{14}). In other words, $\alpha$ satisfies the equation
\be\label{23}
\alpha'=\frac{\sin\alpha}{\sqrt{L^2+\lm\sin^2\alpha}},\quad \alpha(-\infty)=(n-1)\pi,\quad \alpha(\infty)=n\pi.
\ee
On the other hand, with the change of variable $x\mapsto -x$, we flip the problem (\ref{23}) into
\be\label{24}
\alpha'=-\frac{\sin\alpha}{\sqrt{L^2+\lm\sin^2\alpha}},\quad \alpha(-\infty)=n\pi,\quad \alpha(\infty)=(n-1)\pi.
\ee
Hence we may concentrate on (\ref{23}) without loss of generality where $n-1$ is an even integer. Moreover, since (\ref{23}) is invariant under the translation
$\alpha\mapsto \alpha-(n-1)\pi$ (because $n-1$ is even), we may simply consider the integration of the reduced problem
\be\label{25}
\alpha'=\frac{\sin\alpha}{\sqrt{1+\kappa\sin^2\alpha}},\quad \alpha(-\infty)=0,\quad \alpha(\infty)=\pi,
\ee
where we have suppressed the coupling parameters  and rescaled the coordinate variable:
\be 
\frac\lm{L^2}=\kappa,\quad \frac xL\mapsto x.
\ee

To proceed, we first recast the differential equation in (\ref{25}) into
\be \label{27}
\frac{\sqrt{1+\kappa(1-\cos^2\alpha)}}{1-\cos^2\alpha}\dd\cos\alpha=-\dd x.
\ee

With the new variable $u=\cos\alpha$ we consider the integral
\be \label{I}
I=\int\frac{\sqrt{1+\kappa(1-u^2)}}{1-u^2}\,\dd u.
\ee

Introduce the transformation
\be 
\sqrt{1+\kappa(1-u^2)}=\sqrt{1+\kappa}+uv.
\ee
We have the relations
\bea 
u&=&-\frac{2v\sqrt{1+\kappa}}{v^2+\kappa},\\
\sqrt{1+\kappa}+uv&=&-\frac{\sqrt{1+\kappa}(v^2-\kappa)}{v^2+\kappa},\\
1-u^2&=&\frac{(v^2-\kappa)^2-4v^2}{(v^2+\kappa)^2},\\
\dd u&=&\frac{2\sqrt{1+\kappa}(v^2-\kappa)}{(v^2+\kappa)^2}\,\dd v.
\eea
Thus we may carry out an integration of (\ref{I}) as follows, 
\bea\label{II}
I&=&-2(1+\kappa)\int\frac{(v^2-\kappa)^2}{([v^2-\kappa]^2-4v^2)(v^2+\kappa)}\,\dd v\nn\\
&=&\int\left(\frac{(v+1)}{v^2+2v-\kappa}-\frac{(v-1)}{v^2-2v-\kappa}-\frac{2\kappa}{v^2+\kappa}\right)\,\dd v\nn\\
&=&\frac12\ln\left|\frac{v^2+2v-\kappa}{v^2-2v-\kappa}\right|-2\sqrt{\kappa}\arctan\frac v{\sqrt{\kappa}}+c,
\eea
where $c$ is a constant.  Let $x_0\in(-\infty,\infty)$ be such that
\be 
 \alpha(x_0)=\frac\pi2.
\ee
Hence $u(x_0)=v(x_0)=0$. Integrating (\ref{27}), using (\ref{II}), and 
inserting the condition $v(x_0)=0$, we arrive at the result
\be 
2\sqrt{\kappa}\arctan\frac v{\sqrt{\kappa}}-\frac12\ln\left|\frac{v^2+2v-\kappa}{v^2-2v-\kappa}\right|=x-x_0.
\ee
Returning to the original parameters and coordinate variable, we obtain the solution
\be 
\left.\begin{array}{rll}&&2\sqrt{\lm}\arctan\frac{Lv}{\sqrt{\lm}}-\frac L2\ln\left|\frac{L^2(v^2+2v)-\lm}{L^2(v^2-2v)-\lm}\right|=x-x_0,\\
&&\\
&&Lv\cos\alpha=\sqrt{L^2+\lm\sin^2\alpha}-\sqrt{L^2+\lm},\end{array}\right\}
\ee
explicitly, expressed in an implicit function relation.

We note that a similar construction for the kink solitons arising in an ${\cal N}=2$ supersymmetric theory \cite{ABS} with two flavors of quarks in the study of monopole confinement
\cite{SW,MY,ABE,HD,MY2,T,Gr} has been obtained in \cite{CLY}.
\medskip
\medskip

The research of Chen was supported in part by
Henan Basic Science and Frontier Technology Program
Funds under Grant No. 112300410054. The authors thank the referees for some helpful suggestions.


\begin{thebibliography}{99}

\bibitem{D}
G. H. Derrick,
{\em J. Math. Phys.} {\bf5} (1964) 1252. 

\bibitem{JT}
A. Jaffe and C. H. Taubes, {\em Vortices and Monopoles}, Birkh\"{a}user, Boston, 1980.

\bibitem{MS}
N. Manton and P. Sutcliffe, 
{\em Topological Solitons},
Cambridge Monographs on Mathematical Physics, Cambridge U. Press, Cambridge, 2004.

\bibitem{R}
R. Rajaraman, {\em Solitons and Instantons}, North Holland, Amsterdam, 1982. 

\bibitem{Ybook}
Y. Yang, {\em Solitons in Field Theory and Nonlinear Analysis}, Springer-Verlag, New York, 2001.

\bibitem{S}
T. H. R. Skyrme, {\em Proc. Roy. Soc.} A {\bf260} (1961) 127;
{\em ibid} A {\bf262} (1961)
237;
{\em Nucl. Phys.} {\bf31} (1962) 556;
{\em Internat. J. Mod. Phys.} A {\bf3} (1988) 2745.

\bibitem{B}
E. B. Bogomol¡¯nyi, {\em Sov. J. Nucl. Phys.} {\bf24} (1976) 449.

\bibitem{PS}
M. K. Prasad and C. M. Sommerfield, {\em Phys. Rev. Lett.} {\bf35} (1975) 760.

\bibitem{AD}
M. F. Atiyah, V. G. Drinfeld, N. J. Hitchin, and Yu. I. Manin, {\em Phys. Lett.} A {\bf65} (1978) 185.

\bibitem{W}
E. Witten, {\em Phys. Rev. Lett.} {\bf38} (1977) 121.

\bibitem{A}
A. Actor, {\em Rev. Mod. Phys.} {\bf51} (1979) 461.

\bibitem{NS}
C. Nash and S. Sen, {\em Topology and Geometry for Physicists}, Academic, London and New
York, 1983.

\bibitem{M}
N. S. Manton, {\em Commun. Math. Phys.} {\bf111} (1987) 469.

\bibitem{MR}
N. S. Manton and P. J. Ruback, {\em Phys. Lett.} B {\bf181}
(1986) 137.

\bibitem{E}
M. Esteban, {\em Commun.
Math. Phys.} {\bf105} (1986) 571;
Erratum, {\em ibid} {\bf251} (2004)
209.

\bibitem{L}
E. H. Lieb, {\em Proc. Symposia Pure Math.} {\bf54} (1993) 379.

\bibitem{LY}
F. Lin and Y. Yang, {\em  Commun. Math. Phys.} {\bf249} (2004) 273; {\em ibid} {\bf269} (2007) 137.

\bibitem{AW}
C. Adam and A. Wereszczynski, {\em J. High Energy Phys.} {\bf 1308} (2013) 062.

\bibitem{C1}
F. Canfora, Nonlinear superposition law and Skyrme crystals, {\em Phys. Rev.} D, to appear.

\bibitem{C2}
F. Canfora and P. Salgado-Rebolledo, {\em Phys. Rev.} D {\bf87} (2013) 045023.

\bibitem{C3}
F. Canfora and H. Maeda, {\em Phys. Rev.} D {\bf87} (2013) 084049.

\bibitem{ABS}
R. Auzzi, S. Bolognesi, and M. Shifman, {\em Phys. Rev.} D {\bf81} (2010) 085011.

\bibitem{SW}
 N. Seiberg and E. Witten, {\em Nucl. Phys.} B {\bf426} (1994) 19; {\em ibid}
B {\bf430} (1994) 485(E).

\bibitem{MY}
 A. Marshakov and A. Yung, {\em Nucl. Phys.} B {\bf647} (2002) 3.

\bibitem{ABE}
R. Auzzi, S. Bolognesi, J. Evslin, K. Konishi, and
A. Yung, {\em Nucl. Phys.} B {\bf 673} (2003) 187.

\bibitem{HD}
 A. Hanany and D. Tong, {\em J. High Energy Phys.} {\bf07} (2003) 037.

\bibitem{MY2}
 M. Shifman and A. Yung, {\em Rev. Mod. Phys.} {\bf79}  (2007) 1139.

\bibitem{T}
 D. Tong, {\em Ann. Phys.} (Berlin) {\bf324}  (2009) 30.

\bibitem{Gr}
 J. Greensite, {\em An Introduction to the Confinement Problem},
Lecture Notes in Physics {\bf821}, Springer-Verlag, Berlin, 2011.

\bibitem{CLY}
S. Chen, Y. Li, and Y. Yang, {\em Phys. Rev.} D {\bf86} (2012) 085030.

\end{thebibliography}
\end{document}